# Future Study of Dense Superconducting Hydrides at High Pressure

Dong Wang, Yang Ding* and Ho-Kwang Mao

*Center for High Pressure Science and Technology Advanced Research, Beijing 100094, China.*

*\*Correspondence: yang.ding@hpstar.ac.cn*

**Abstract:** The discovery of a record high superconducting transition temperature ($T_c$) of 288 K in a pressurized hydride inspires new hope to realize ambient-condition superconductivity. Here, we give a perspective on the theoretical and experimental studies of hydride superconductivity. Predictions based on the BCS–Eliashberg–Midgal theory with the aid of density functional theory have been playing a leading role in the research and guiding the experimental realizations. To date, about twenty hydrides experiments have been reported to exhibit high-$T_c$ superconductivity and their $T_c$ agree well with the predicted values. However, there are still some controversies existing between the predictions and experiments, such as no significant transition temperature broadening observed in the magnetic field, the experimental electron-phonon coupling beyond the Eliashberg–Midgal limit, and the energy dependence of density of states around the Fermi level. To investigate these controversies and the origin of the highest $T_c$ in hydrides, key experiments are required to determine the structure, bonding, and vibrational properties associated with H atoms in these hydrides.

## 1. Introduction

Since the discovery of the superconductivity of mercury at 4.2 K by H. K. Onnes in 1911 [1], scientists have been relentlessly hunting for superconducting materials with higher $T_c$ values. There are several 'rushes' in the research history of high-$T_c$ superconductors that were marked by the discoveries of cuprates ($T_c$ of 133 K at ambient pressure [2] and 164 K at high pressure [3]), iron based superconductors ($T_c$ of above 50 K [4,5]), and $MgB_2$ with $T_c$ of 39 K [6].

In 2004, Ashcroft predicted that compressed hydrogen-rich hydrides are good candidates for high-$T_c$ superconductors due to chemical "pre-compression" effects [7]. This idea has driven subsequent efforts on studying the hydrides, eventually leading to the discovery of superconductivity above 200 K, first in $H_3S$ ($T_c$ of 203 K) [8] and more recently in $LaH_{10}$ ($T_c$ up to 260 K) [9,10] and a carbonaceous sulfur hydride (CSH) ($T_c$ of 288 K) [11]. Since 2015, superconductivity has been reported in the following compounds at high pressure: $PH_x$ above 100 K under 207 GPa [12], $YH_x$ at 243 K above 200 GPa [13,14], $ThH_x$ at 161 K under 175 GPa [15], $PrH_x$ at 9 K under 130 GPa [16], $LaYH_x$ at 253 K under 183 GPa [17], $CeH_x$ at 115–120 K under 95 GPa [18], $SnH_x$ at 70 K under 200 GPa [19], $BaH_x$ around 20 K under 140 GPa [20], $CaH_x$ at 215 K under 172 GPa [21], $ScH_x$ at 22.4 K under 156 GPa, and $LuH_x$ at 15 K under 128 GPa [22]. The experiment conditions and results are summarized in the Table 1. In these experiments, electric resistance and powder X-ray diffraction are commonly performed. However, due to the very weak scattering power and background from the pressure vessels, the position of H in the hydrides is unable to be determined by experiments.

**Table 1.** The experiment conditions and results of superconducting hydrides.

| Hydrides | Starting Materials | Laser Heating | Electric Resistance | XRD | Isotope Effects | Magnetization | $T_c$/Pressure | Ref. |
|---|---|---|---|---|---|---|---|---|
| $PH_x$ | Liquid $PH_3$ | No | $R(T_c,\text{onset}) = \sim 5\ \Omega$ Zero resistance | — | — | — | 103 K/207 GPa | [12] |
| $YH_x$ | $Y + H_2/D_2$ $YH_3/YD_3 + H_2/D_2$ $YH_3 + AB$ | 2000(10) K | $R(T_c,\text{onset}) = \sim 0.37\ \Omega$ Zero resistance | Yes | Yes | — | 243 K/201 GPa | [13] |
| $ThH_x$ | Th + AB | 179 GPa, 1.064 nm laser, power: 35–40 W | $R(T_c,\text{onset}) = \sim 0.081\ \Omega$ Zero resistance | Yes | — | — | 161 K/175 GPa | [15] |
| $PrH_x$ | Pr + AB | 1650 K@115 GPa | $R(T_c,\text{onset}) = \sim 0.9\ \Omega$ | Yes | — | — | ~9 K/130 GPa | [16] |
| $LaYH_x$ | LaY alloys + AB | 2000 K@170–196 GPa | $R(T_c,\text{onset}) = \sim 0.392\ \Omega$ Zero resistance | Yes | — | — | 253 K/183 GPa | [17] |
| $LaH_x$ | $La + H_2$ $La + D_2$ | 1500 K@145 GPa | $R(T_c,\text{onset}) = \sim 0.95\ \Omega$ Zero resistance | Yes | Yes | — | 250 K/170 GPa | [9] |
| | La + AB | 2000 K@~180 GPa | $R(T_c,\text{onset}) = \sim 0.92\ \Omega$ Zero resistance | Yes | — | — | 260 K/188 GPa | [10] |
| $SnH_x$ | Sn + AB | 1700 K@200 GPa | $R(T_c,\text{onset}) = \sim 0.52\ \Omega$, | Yes | — | — | ~70 K/200 GPa | [19] |
| CSH | $C + S + H_2$ | 4.0 GPa 532-nm laser light, power: 10–25 mW. | $R(T_c,\text{onset}) = \sim 1.66\ \Omega$ Zero resistance | Yes | — | Yes | 287.7 K/267 GPa | [11] |
| $H_xS$ | Liquid $H_2S$ or $D_2S$ | No | Zero resistance | Yes | Yes | Yes | 203 K/155 GPa [a] | [8] |
| $CaH_x$ | Ca foil + AB | 2000 K@160–190 GPa | $R(T_c,\text{onset}) = \sim 0.97\ \Omega$ Zero resistance | Yes | — | — | 215 K/172 GPa | [21] |
| $CeH_x$ | Ce + AB Ce + $ND_3BH_3$ Ce + $ND_3BD_3$ Ce + $D_2$ | 1500 K@90–137 GPa | Zero resistance | Yes | Yes | — | 115 K/95 GPa | [18] |
| $BaH_x$ | Ba + AB | 1600 K@90 GPa | $R(T_c,\text{onset}) = \sim 1.12\ \Omega$ | Yes | — | — | 20 K/140 GPa | [20] |
| $ScH_x$ | Sc + AB | 1500 K | $R(T_c,\text{onset}) = 0.26\ \Omega$ | Yes | — | — | 22.4 K/156 GPa | [22] |
| $LuH_x$ | Lu + AB | 1500 K/110 GPa | $R(T_c,\text{onset}) = 0.05\ \Omega$ Zero resistance | Yes | — | — | 15 K/128 GPa | [22] |

Notes: AB: $NH_3BH_3$; "—"means this term is not mentioned in corresponding literature; "Yes" means this measurement was mentioned to characterize the hydrides; [a] value comes from Magnetic susceptibility measurement.

To demonstrate the relationship between $T_c$ and pressure, we plot the figure of merit $S$, which is an index designed [23,24] to compare experimental reported superconductors. The $S$ value of $MgB_2$ is 1, which serves as a benchmark. The higher the merit index $S$, the higher superconducting $T_c$ of hydride under comparable high pressure. Amongst these, $CeH_x$, to date, has the highest $S$ value as 1.12 below 100 GPa (Figure 1). The isotope effects were also experimentally observed in $D_2S$ [8], $YD_6$ [14], $LaD_{10}$ [9], and CeD₉ [18]. These experimental results (see Figure 1) have confirmed theoretical predictions for the basic structures and superconducting properties [25–40], thereby starting a new era for exploring the ambient condition superconductivity.

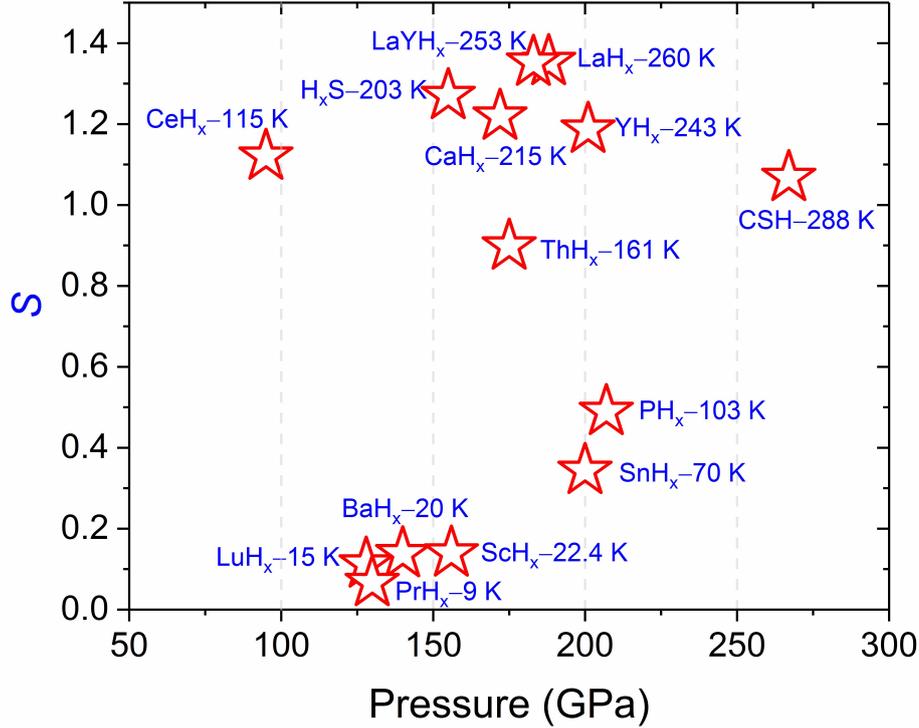

**Figure 1.** The superconducting hydrides reported from experiments. Merit index $S$, calibrated with $T_c$ of MgB$_2$ 39 K, is estimated as $T_c/((T_{c,MgB_2})^2 + \text{Pressure}^2)^{1/2}$ that is introduced in Reference [23], $T_{c,MgB_2}$ is the superconductivity transition temperature (39 K) of conventional superconductor MgB$_2$, $T_c$ represents the superconductivity transition temperature ($T_c$) of a hydride under a certain pressure ($P$).

Currently, predictions based on the BCS–Eliashberg–Midgal theory [41–43] with the aid of the first principle density functional theory (DFT) calculations [44,45] have already contributed a large amount of computational data on binary hydrides [25–37], and more works are being rapidly extended to the ternary, quadratic, and non-hydrogen systems. Recently several new hydrides with high $T_c$ values are predicted, including a binary hydride $NaH_6$ with a $T_c$ of 267 K within 100 GPa [23], a ternary hydride $Li_2MgH_{16}$ of a $T_c$ of 473 K at 250 GPa [46], and another ternary hydride $LaBH_8$ with a $T_c$ of 126 K at 50 GPa [47], which are driving more experimental discoveries in the future.

## 2. Discussions

### 2.1. Unexplained Anomalous Superconducting Behaviors in Hydrides

According to the established understanding of superconductivity [48], both typical conventional and unconventional superconductors are universally characterized by three parameters: the London penetration depth $\lambda_L$, the Peppard coherence length $\xi$, and the electron mean free path. Type I superconductors with $\lambda_L < \xi$ are a limited set of materials, mostly pure forms of elements, which are all considered as conventional superconductors [49], while most superconducting materials, including both conventional and unconventional, are type II with $\xi < \lambda_L$ or strong type II have $\xi \ll \lambda_L$ superconductors. According to Ginzburg–Landau (GL) theory, strong type II superconductors commonly show a broadening of transition temperature as $\Delta T/T_c \sim 0.02$, which increases with increasing applied magnetic field. For instance, in cuprate it finds $\Delta T/T_c \propto H^{2/3}$ [50]. Intriguingly, J. E. Hirsch [51–53] points out that the $\Delta T/T_c$ of hydrides shows no significant broadening in the increasing magnetic field, as examples in Figure 2 shown. Even $\Delta T/T_c$ of $H_3S$ without magnetic field are regarded to be too small according to GL theory [11,52,53]. Therefore, although the superconducting hydrides have been accepted as typical BCS conventional superconductors, their most notable anomalous superconducting behavior remains unexplained by the classical GL theory.

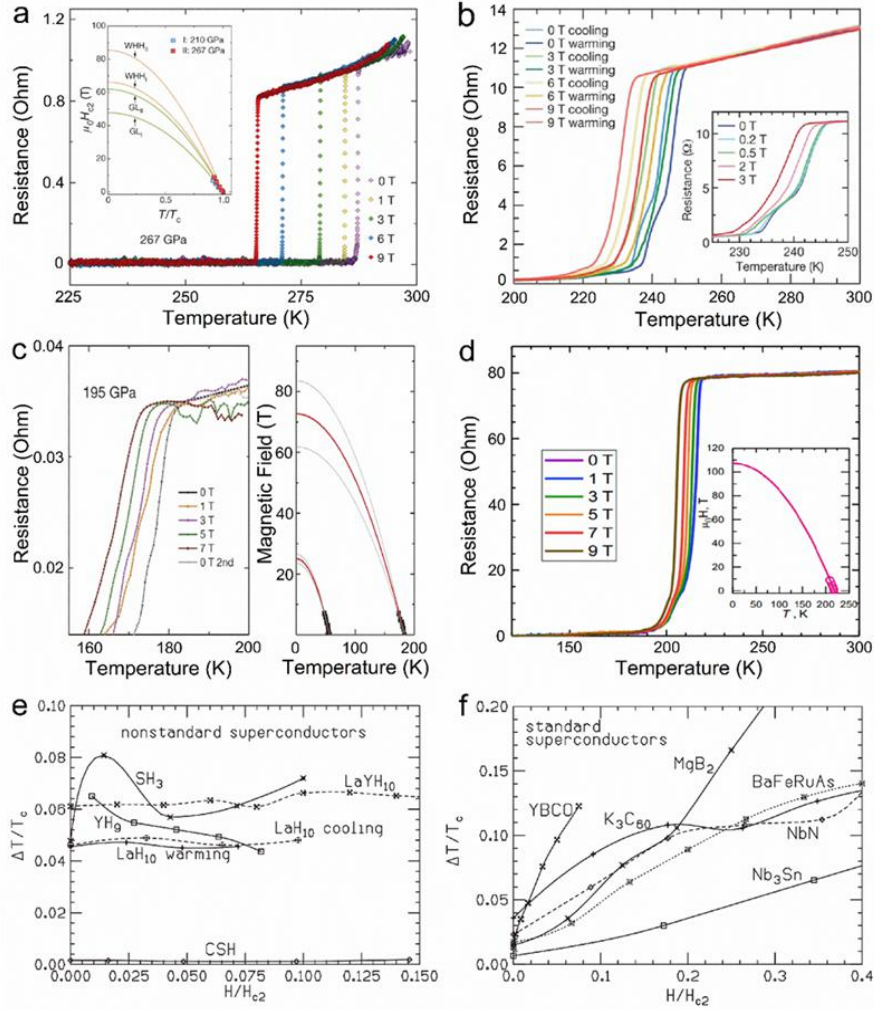

**Figure 2.** Resistive transition values for (**a**) CHS [11], (**b**) LaH$_{10}$ [9], (**c**) H$_3$S [8] and (**d**) YH$_9$ [13] in a magnetic field ranging from 0 to 9 T, which, in contrast to most typical type II superconductors (**f**) [52], shows no appreciable broadening ($\Delta T/T_c$) with increasing of magnetic field (**e**) [52].

The contradiction between the successful and accurate prediction of $T_c$ in many high-pressure hydrides, and the fact that they do not follow the classical GL theory, requires further investigations in which the utmost urgent thing is to experimentally prove whether the high-pressure hydrides are superconductors or not by confirming if the key Meissner effect coexists with zero resistance. To date, a total of about seven magnetization measurements have been reported to observe the Meissner effect in CSH [11], LaH$_x$ [38,54], and H$_x$S [8,54–56] with magnetic susceptibility and nuclear resonant scattering (Table 2). However, the validity of these results has been strongly questioned by J. E. Hirsch [57–59] and the question remains open.

**Table 2.** The list of reported magnetization measurements.

| Hydrides | Methods | Year | Ref. |
|:---:|:---:|:---:|:---:|
| H$_x$S | AC magnetic susceptibility measurements | 2015 | [8] |
| H$_x$S | nuclear resonant scattering | 2016 | [55] |
| H$_x$S | AC magnetic susceptibility measurements | 2019 | [56] |
| LaH$_x$ | AC magnetic susceptibility measurements | 2020 | [38] |
| CSH | AC magnetic susceptibility measurements | 2020 | [11] |
| H$_x$S/LaH$_x$ | accurate magnetometry measurements | 2021 | [54] |

This anomalous behavior of hydrides can be caused by several possible reasons: (1) the pressure increases during cooling [60], which suppresses the broadening of $T_c$. This phenomenon is commonly observed in low temperature diamond anvil experiments. To solve this problem, it requires minimizing the variation of the pressure values during the cooling process or accurately determining how the pressure changes with temperature [60]. Then a reliable temperature dependence of resistance curve could be measured to determine the broadening of $T_c$; (2) the significant resistance drop might be caused by other phase transitions rather than superconducting transition [52], such as metallization phase transitions, structural or magnetic phase transitions. These possibilities can be verified by investigating if the changes in crystal structure, phonon structure, electronic structure, or magnetic structure concur with the resistance drops. More importantly, the Meissner effects must be conclusively confirmed to concur with the resistance drops; (3) it is a new type of superconductivity. That means not only GL theory but also BCS theory is inapplicable to this type of superconductivity, and new mechanisms need to be explored in the first priority for hydrides.

## 2.2. How to Understand the Origin of the Highest $T_c$ in Hydrides

The BCS theory accounts for why hydrides could have high-$T_c$ values, but is unable to describe material-related properties, such as why $LaH_{10}$ and CSH have the highest $T_c$ values but other hydrides do not. Is there any other hydride having even higher $T_c$ values? The answers to these questions are still unclear but fundamentally important for searching the higher $T_c$ superconductors.

Marvin Cohen has long believed that the secret for increasing superconducting temperatures resides in covalent bonds; he asserts that these new compounds allow for testing of this idea [61]. Actually in 1971, J. Gilman already predicted the possibility of making a new form of hydrogen in a metallic state through the preparation of a covalent compound $LiH_2F$ under pressure [62]. Quan and Pickett proposed the metallization of the covalent bond is the key driving force for high-$T_c$ in $MgB_2$ [63]. Covalent metals such as $MgB_2$ are rare at ambient pressure but may be formed and become stabilized at high pressures. Indeed, it is noteworthy that so far all systems with $T_c > 200$ K under pressure are hydrogen rich compounds, typically the covalent $H_3S$ [8], $LaH_{10}$ [9] and CSH [64,65] systems. Besides, covalent hydrogen-rich organic-derived materials are another class of high-$T_c$ materials that combine the advantages of covalent metals and metal superhydrides. A common feature for all these classes of materials is the existence of covalent bonds, which probably implies that strong-covalent bonding is the key to driving the high-$T_c$ superconductivity in hydrides, as shown in Figure 3a–d. In addition, van Hove singularity around the Fermi level is also observed in $H_3S$ [25,63,66,67] and $LaH_{10}$ [68] from calculations, which shows remarkable effects on $T_c$ and thus is regarded as a possible origin for high $T_c$ (see Figure 3e,f). Recently, Coulomb effects [69], Lifshitz transition [70], multiband pairing [71] and anharmonicity [27,72–76] are also proposed to account for the high-$T_c$ superconductivity in the dense hydrides. Therefore, more efforts are demanded on investigating the origin of material-dependent high-$T_c$ in hydrides, especially identifying and confirming the role of metallic covalent bonds in the high-$T_c$ superconducting hydrides.

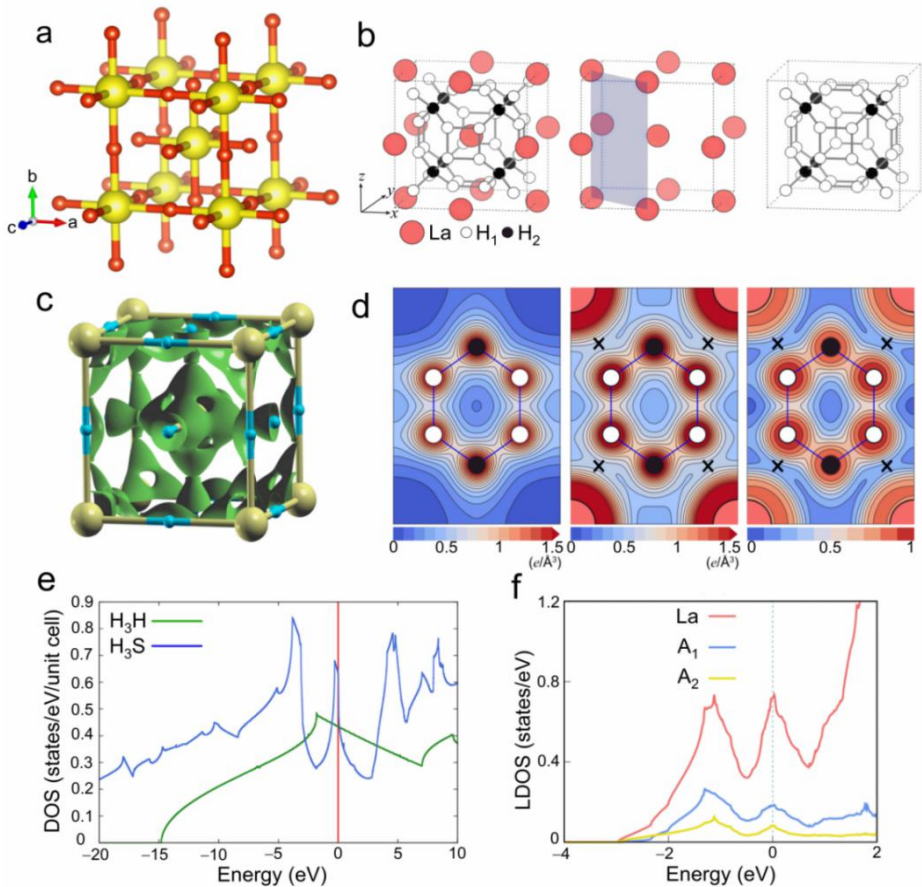

**Figure 3.** Predicted crystal structure models for (**a**) $H_3S$ (red = H atom and yellow = sulfur atom) and (**b**) $LaH_{10}$. (**c**) Calculated isosurface of the charge density obtained from states within the energy range of ± 1 eV (yellow = sulfur atom, blue = H atom), which demonstrates the covalent bonding forming between sulfur atoms and H atoms with revealed by in $H_3S$. (**d**) Calculated total charge densities of $LaH_{10}$ with the contour

spacing of 0.1 e/Å³ (left), the charge connection between La and H atoms around the point marked "×" represents the metallic covalent character of the La−H bond (mid), and the ELF of LaH$_{10}$ with the contour spacing of 0.1 (right). (**e**) Calculated density of states of H$_3$S compared to that of H$_3$H at the same volume. The peak at Fermi level arises van Hove singularities. (**f**) Calculated DOS of LaH$_{10}$ at 300 GPa. The peak at Fermi level is associated with van Hove singularity. The data in (**a**) (**c**) and (**e**) for H$_3$S are cited from Reference [63] and the data in (**b**) (**d**) and (**f**) for LaH$_{10}$ are cited from Reference [77].

### 2.3. The Limits of the BCS-Eliashberg Theory

It is shown that electron-phonon coupling constant $\lambda$ for many hydrides is typically predicted to above 2 and some even reaches 5.8 in CaH$_6$ [23]. Provided BCS–Eliashberg-Migdal theory usually accounts for the universal behavior of conventional superconductors with $\lambda < 1.5$–2, the validity of the theory for strongly coupling hydrides is still under debate [78–80]. In addition, anharmonic effects and energy dependent electronic structure also play significant roles in the high-$T_c$ the superconductivity, which are neglected in the theory and requires particular attention in the future predictions.

(1) The Migdal-Eliashberg (ME) theory [41,42], in which vertex corrections are neglected for simplification, usually describes electron-phonon coupling effects accurately for conventional superconductors. As the electron-electron repulsion in the theory is usually approximated by an empirical parameter $\mu^*$ (mostly considered to be ~0.1) to reproduce the experimental $T_c$ [43], this approach becomes less accurate in the limit of sizeable electron–phonon coupling or in the case of systems with strongly anisotropic electronic properties [68,78–80]. It is more appropriate to use other methods based on a perturbative Green's function approach such as the full ab-initio Migdal–Eliashberg approach or SCDFT method [44,45,68]. The latter recently has been applied to reproduce the $T_c$ of $H_3S$, in good agreement with the experiment [44].

(2) Neglecting the energy dependence of density of states (DOS) around the Fermi level in the Migdal–Eliashberg theory for simplification may overlook some peculiar energy dependent electronic structures occurring in hydrides (i.e van Hove singularity in $H_3S$ [63,66,81,82] and $LaH_{10}$ [68], Fermiology due to Lifshitz transitions [70]) around Fermi level, which may suppress the $T_c$. For instance, predictions beyond the constant DOS approximation, by explicitly considering the electronic structure around the Fermi level in $H_3S$, show the constant DOS approximation employed, to date, overestimates $T_c$ by ~ 60 K, or underestimates by ~ 10 K when the energy dependence of DOS are present or absent near the Fermi level [66], respectively.

(3) Because of the low mass of hydrogen and its large quantum fluctuations from equilibrium, substantial anharmonic corrections to $T_c$ have been predicted in some superconducting hydrides and phases of hydrogen [27,72–76]. For instance, PtH at 100 GPa shows strong anharmonic hardening of the phonon energies, which suppresses the $T_c$ by over an order of magnitude [73]. Anharmonic effects are also predicted to lead to an inverse isotope effect in superconducting palladium hydride [74] and cause the value of $T_c$ falls 22% from 250 K to 194 K in $H_3S$ [27]. Thus, it becomes urgent to extensively develop an understanding of the anharmonic effects on $T_c$ of the superconducting dense hydrides.

In addition, the BCS theory also faces some other unexplained experimental phenomena, such as the reversible phase transition between the normal and superconducting phases in the H-T plane (for type I superconductors); the electron mass in the London magnetic field should be twice of the free electron mass ($2m$) rather than twice of an effective mass ($2m^*$) as predicted. These phenomena can now be explained with a new theory which attributes the superfluid to the nonlinear Berry connection emerging from the many-body wave function connection [83–85], which might also be applicable to the superconductivity of hydrides.

*2.4. More Experiments Are Demanded*

Due to the complexity and challenges of high-pressure experiments, the high-pressure measurements of hydrides is only limited to <20 groups, which stays far behind theoretical predictions (more than 5000), and the experiment results are largely outnumbered by those theoretical predictions [68]. The experiments are not only performed to confirm and the benchmark of theoretical predictions, but it is more importantly to reveal breakthrough discoveries that may be overlooked by theories. Thus, in addition to routine resistance measurements, more experimental efforts should be devoted to developing new high-pressure techniques facilitating the following measurements on the superconducting hydrides.

(1) Crystal structure determination of hydrides. Crystal structure is the most fundamental information, however, to date, the positions of H atoms in hydrides remain undetermined with conventional X-ray methods due to their weak scattering power. Consequently, all hydrogen network and the nature of bonding predicted from theory have never been confirmed. However, owing to the new technology developments of high pressure X-ray diffraction beamline for diamond anvil cell [86], recently a successful unit cell parameters determination of the phase IV of hydrogen at 200 GPa with synchrotron X-ray by Ji et al. [87,88] shows promise for conquering the problems in the future. Powder neutron diffraction has been applied at high pressure study up to 90 GPa and could be an ideal probe to study the H structures [89].

(2) The magnetic responses of superconducting hydrides. To date, only seven magnetization measurements have been reported: AC magnetic susceptibility measurements for CSH [11], three for $H_xS$ [8,54,56], $LaH_x$ [38,54], the nuclear resonant scattering measurements for $H_3S$ [55]. However, the magnetic signal from samples are commonly complicated by the noise from backgrounds [57–59], which leaves the reported Meissner effects in debate [57–59]. To increase the required signal-to-background ratio required the development of a new high-pressure technique, such as specially designed miniature nonmagnetic DAC cells made of Cu-Ti alloy is needed to accommodate in a SQUID magnetometer [90].

(3) Electronic and vibrational properties of hydrides. The electron pairing in hydrides are mediated by electron-phonon coupling, which is essentially associated with interactions between the electronic states near Fermi level and the high frequencies phonons of H atoms. The information on the electronic and vibrational properties of H is crucial for understanding the mechanisms underlying the material-dependent high $T_c$ in hydrides. Experimentally, the electronic structure of H atoms (and host atoms) can be probed with X-ray Raman [91] and nuclear magnetic resonance spectroscopy [92]. The vibrational properties of hydrides can be obtained from the phonon dispersion, which can be measured with the high resolution (meV)-energy resolution inelastic X-ray scattering. The zone-center vibrational optical modes and superconducting gap can be also studied with or Raman and Infrared spectroscopes [93].

Combing the results from (1)–(3), it is also possible to determine (a) if the anomalous resistive behavior hydrides is originated from superconducting transition or other phase transitions or errors from experiments; and (b) the anharmonic effects on the $T_c$.

## 2.5. Perspective

The exploration of room temperature superconductors has been a coveted goal of many scientists. The progress in the last 30 years has been tremendous. The discovery of new superconductors, including cuprates, heavy fermion superconductors, organic superconductors, iron-based superconductors, two-dimensional superconductors, topological superconductors, nickelates, and now dense superhydrides, has not only provided important topics for physics and materials science but also opened up new research fields. The research of superconductivity has gone through the London equation [94], Landau-Ginzberg theory [95], and BCS theory [96]. The Meissner effect, as a result of spontaneous symmetry breaking, is a manifestation of the Anderson–Higgs mechanism [97–99]. The BCS theory is one of the most important theoretical advances after the establishment of quantum mechanics and has been successfully applied to the conventional superconductors. However, the mechanism of high-temperature superconductivity remains as a grand-challenge in physics and a unified theory remains elusive. To date, the existence of d-wave symmetry and pseudo-energy gaps in copper-oxygen is well established [100–102]. Studies of iron-based superconductors, organic superconductors, heavy fermion superconductors, two-dimensional superconductors, nickelates, and topological superconductors have also shown the existence of multiple cooperative and competing orders is a universal phenomenon [103–108]. These results provide a base for developing ultimate theories in the future.

The discovery of the room-temperature superconductivity in dense superhydrides at high pressure is unexpected but promises a completely new research direction for the realization of high temperature (or room temperature) superconductivity at ambient pressure. Due to the complicated high-pressure environments, the current research is limited with experimental techniques because many advance techniques such as scanning tunneling electron microscope (STEM), angular resolved photoelectron emission spectroscopy (ARPES), neutron inelastic scattering (NIS) are inapplicable at high pressure conditions, and also faces some challenges to obtain high-quantity signals from samples. However, in contrast to other superconducting materials, the experimentally measured superconducting temperature in hydrides commonly shows an excellent agreement with the theoretical predicted ones. This implies theoretical calculations might be able to play a leading and active role in the research of hydride superconductivity. Apart from aforementioned unexplained and debated experimental results, several possible research directions could be prospected: (1) developing reliable methods to accurately predict the superconducting $T_c$; (2) exploring the ternary and quaternary hydrides by increases the number of possible structures and thereby the more choices of low-pressure-high-$T_c$ hydrides; (3) replacing H with other light atoms, such as B and C. Materials made of boron and carbon also form covalent bonds and have phonons suitable for high-temperature superconductivity; (4) developing novel techniques to enable more key measurements and advance the research of superconducting hydrides. The ultimate goal is to achieve the room temperature superconductivity at low or ambient pressure.

## 3. Conclusions

Since 2004, many hydrides have been reported to show (nearly) zero resistance at high pressure with the transition temperature $T_c$ values close to the ones predicted with BCS–Eliashberg Theory. With magnetization and isotopic effects measurements, these zero-resistance hydrides have been widely accepted as typical BCS conventional superconductors, even though their anomalous superconducting behavior remained unexplained with the classical GL theory. The contradiction between the successful prediction of $T_c$ and unsuccessful application of the classical GL theory becomes an utmost urgent question to be investigated.

To search for or make the higher $T_c$ superconductors, more research is demanded to investigate the key driving force for the high $T_c$ superconductivity in hydrides, including the metallization of the strong covalent bond, van Hove singularity around the Fermi level, Coulomb effects, Lifshitz transition, multiband pairing, and anharmonicity.

The validity of the BCS–Eliashberg–Migdal theory for strongly coupling hydrides is still in debate, which may limit its application in prediction of the $T_c$ in dense hydrides. In addition, anharmonic effects and energy dependent electronic structure may also play significant roles in the high-$T_c$ the superconductivity, which requires particular attention in the predictions.

The discovery of the room-temperature superconductivity in dense superhydrides at high pressure promises a new research direction for the realization of high temperature (or room temperature) superconductivity at ambient pressure. Some key experimental results remained to be confirmed, and thus more efforts need to be devoted to studying the structure, bonding, and vibrational properties of H atoms in hydrides. It is also expected that the breakthrough discoveries from experiments could lead to revolutions in theories [109].

**Funding:** Y. Ding is grateful for support from the National Key Research and Development Program of China 2018YFA0305703, Science Challenge Project, No TZ2016001, and the National Natural Science Foundation of China (NSFC):11874075. H.K.M. acknowledges supports from the National Natural Science Foundation of China Grant No. U1530402 and U1930401.

**Conflicts of Interest:** The authors declare no conflict of interest.